\documentclass[aps,preprint]{revtex4}
\usepackage{graphicx}
\usepackage[scanall]{psfrag}
\usepackage{amssymb}

\begin{document}

%
%
%
%
\def\oti{{\otimes}}
\def\lb{ \left[ }
\def\rb{ \right]  }
\def\tilde{\widetilde}
\def\bar{\overline}
\def\hat{\widehat}
\def\*{\star}
\def\[{\left[}
\def\]{\right]}
\def\({\left(}		\def\BL{\Bigr(}
\def\){\right)}		\def\BR{\Bigr)}
	\def\BBL{\lb}
	\def\BBR{\rb}
%
%
\def\zb{{\bar{z} }}
\def\zbar{{\bar{z} }}
\def\frac#1#2{{#1 \over #2}}
\def\inv#1{{1 \over #1}}
\def\half{{1 \over 2}}
\def\d{\partial}
\def\der#1{{\partial \over \partial #1}}
\def\dd#1#2{{\partial #1 \over \partial #2}}
\def\vev#1{\langle #1 \rangle}
\def\ket#1{ | #1 \rangle}
\def\rvac{\hbox{$\vert 0\rangle$}}
\def\lvac{\hbox{$\langle 0 \vert $}}
\def\2pi{\hbox{$2\pi i$}}
\def\e#1{{\rm e}^{^{\scriptstyle #1}}}
\def\grad#1{\,\nabla\!_{{#1}}\,}
\def\dsl{\raise.15ex\hbox{/}\kern-.57em\partial}
\def\Dsl{\,\raise.15ex\hbox{/}\mkern-.13.5mu D}
%
%
\def\ga{\gamma}		\def\Ga{\Gamma}
\def\be{\beta}
\def\al{\alpha}
\def\ep{\epsilon}
\def\vep{\varepsilon}
\def\la{\lambda}	\def\La{\Lambda}
\def\de{\delta}		\def\De{\Delta}
\def\om{\omega}		\def\Om{\Omega}
\def\sig{\sigma}	\def\Sig{\Sigma}
\def\vphi{\varphi}

%
%
\def\CA{{\cal A}}	\def\CB{{\cal B}}	\def\CC{{\cal C}}
\def\CD{{\cal D}}	\def\CE{{\cal E}}	\def\CF{{\cal F}}
\def\CG{{\cal G}}	\def\CH{{\cal H}}	\def\CI{{\cal J}}
\def\CJ{{\cal J}}	\def\CK{{\cal K}}	\def\CL{{\cal L}}
\def\CM{{\cal M}}	\def\CN{{\cal N}}	\def\CO{{\cal O}}
\def\CP{{\cal P}}	\def\CQ{{\cal Q}}	\def\CR{{\cal R}}
\def\CS{{\cal S}}	\def\CT{{\cal T}}	\def\CU{{\cal U}}
\def\CV{{\cal V}}	\def\CW{{\cal W}}	\def\CX{{\cal X}}
\def\CY{{\cal Y}}	\def\CZ{{\cal Z}}

\def\rvac{\hbox{$\vert 0\rangle$}}
\def\lvac{\hbox{$\langle 0 \vert $}}
\def\comm#1#2{ \BBL\ #1\ ,\ #2 \BBR }
\def\2pi{\hbox{$2\pi i$}}
\def\e#1{{\rm e}^{^{\textstyle #1}}}
\def\grad#1{\,\nabla\!_{{#1}}\,}
\def\dsl{\raise.15ex\hbox{/}\kern-.57em\partial}
\def\Dsl{\,\raise.15ex\hbox{/}\mkern-.13.5mu D}
%
%
%
\font\numbers=cmss12
\font\upright=cmu10 scaled\magstep1
\def\stroke{\vrule height8pt width0.4pt depth-0.1pt}
\def\topfleck{\vrule height8pt width0.5pt depth-5.9pt}
\def\botfleck{\vrule height2pt width0.5pt depth0.1pt}
\def\Zmath{\vcenter{\hbox{\numbers\rlap{\rlap{Z}\kern
0.8pt\topfleck}\kern 2.2pt
                   \rlap Z\kern 6pt\botfleck\kern 1pt}}}
\def\Qmath{\vcenter{\hbox{\upright\rlap{\rlap{Q}\kern
                   3.8pt\stroke}\phantom{Q}}}}
\def\Nmath{\vcenter{\hbox{\upright\rlap{I}\kern 1.7pt N}}}
\def\Cmath{\vcenter{\hbox{\upright\rlap{\rlap{C}\kern
                   3.8pt\stroke}\phantom{C}}}}
\def\Rmath{\vcenter{\hbox{\upright\rlap{I}\kern 1.7pt R}}}
\def\Z{\ifmmode\Zmath\else$\Zmath$\fi}
\def\Q{\ifmmode\Qmath\else$\Qmath$\fi}
\def\N{\ifmmode\Nmath\else$\Nmath$\fi}
\def\C{\ifmmode\Cmath\else$\Cmath$\fi}
\def\R{\ifmmode\Rmath\else$\Rmath$\fi}

\def\barray{\begin{eqnarray}}
\def\earray{\end{eqnarray}}
\def\beq{\begin{equation}}
\def\eeq{\end{equation}}

\def\no{\noindent}

\def\pmb#1{\setbox0=\hbox{#1}%
\kern-.025em\copy0\kern-\wd0
\kern.05em\copy0\kern-\wd0
\kern-.025em\raise.0433em\box0 }

\def\smallsqrtk{{\scriptstyle \sqrt{k}}}
\def\smallsqrttwo{{ \scriptstyle \sqrt{2}}}

\def\gl{gl(1|1)}
\def\glhat{gl(1|1)}
\def\glN{gl(N|N)}
\def\osp2N{osp(2N|2N)}

\def\psibar{\bar{\psi}}
\def\betabar{\bar{\beta}}

\def\osp{osp(2|2)}
\def\osphat{\hat{osp(2|2)}}
\def\Shat{\hat{S}}

\def\bfphi{\pmb{$\phi$}}
\def\bfchi{\pmb{$\chi$}}

\def\phibar{\bar{\phi}}
\def\chibar{\bar{\chi}}
\def\sqrtk{{\sqrt{k}}}
\def\rep{{\it r}}
\def\repbar{ {\bar{\it r}} }

\def\glrep#1#2{{\langle #1, #2 \rangle}}
\def\hone#1{{\langle #1\rangle_{(1)}}}
\def\hfour#1{{\langle #1 \rangle_{(4)}}}
\def\Deltahj#1#2{{\Delta_{\glrep{#1}{#2}}}}
\def\chitilde{\tilde{\chi}}

\def\Deltachi{ \Delta^{(\chi)}}

\def\Vbos#1#2{\CV^\phi_{#1,#2}}
\def\V#1#2{V_{\glrep{#1}{#2}}}

\def\muhalf{\mu_{1/2}}
\def\sigmahalf{ \sigma_{1/2}}

\def\Ncopy{{\rm N-copy}}
\def\suN{su(N)}

\def\free{{\rm free}}

\def\Deltamin{\Delta^{\rm (min)}}

\def\bfPhi{{\bf \Phi}}
\def\bfVbos#1#2{{\bf V}^\phi_{#1,#2}}
\def\bfell{\pmb{$\ell$}}
\def\Hbar{\bar{H}}
\def\Jbar{\bar{J}}
\def\repbar{{\bar{\rep}}}
\def\tbar{\bar{t}}
\def\Vbar#1#2{\bar{V}_{\glrep{#1}{#2}}}
\def\bfmu{\pmb{\mu}}
\def\bfsigma{\pmb{\sigma}}
\def\bfPhihj#1#2{\bfPhi_{\glrep{#1}{#2}}}

\def\qvec{\vec{q}}
\def\phivec{\vec{\phi}}

\def\osprep#1#2{ [ #1, #2 ]^{osp}}
\def\surep#1#2{ [ #1, #2]^{su}}

\def\sqrttwo{\sqrt{2}}

\def\eight{[8]^{osp}}

\def\alphavec{\vec{\alpha}}
\def\sigmavec{\vec{\sigma}}
\def\Avec{\vec{A}}

\def\Psibar{\bar{\Psi}}

\def\lie{{\bf g}}
\def\liehat{\hat{{\bf g}}}

\def\texthalf{{\textstyle \inv{2}}}

\def\Gammahat{\hat{\Gamma}}

\def\rme{{\rm e}}

\def\n{\noindent}

\title{Critical points of $2d$ disordered Dirac fermions: 
the Quantum Hall Transitions revisited}
\author{Andr\'e  LeClair}
\affiliation{Newman Laboratory, Cornell University, Ithaca, NY} 
\date{October 2007, Revised May 2008}

\bigskip\bigskip\bigskip\bigskip\bigskip\bigskip\bigskip\bigskip

\begin{abstract}

We propose a resolution of  the  renormalization group flow
for the disordered Dirac fermion theories describing
the quantum Hall transition (QHT)  and spin Quantum Hall transition
(SQHT), which previously 
revealed  no  perturbative  fixed points at 1-loop and higher.  The approach
involves carrying out the flow in 2 stages,  the first stage 
utilizing a new form of super spin-charge separation to flow
to $gl(1|1)_N$ and $osp(2|2)_{-2N}$ supercurrent algebra
theories,  where $N$ is the number of copies.  This fixed point
breaks the copy symmetry.   In the second
stage,  additional forms of disorder are incorporated as
dimension zero logarithmic operators, and the resulting
actions have explicit forms  in terms of two scalar fields and a
symplectic fermion.   Multi-fractal exponents are computed
with the result  $q(1-q)/4$ and $q(1-q)/8$ for the QHT and SQHT 
respectively, in agreement with numerical estimates.

\end{abstract}

\maketitle

\section{Introduction}

Disordered Dirac fermions in $2+1$ dimensions have many
important applications in condensed matter physics.   
They are theoretically interesting since they can 
 represent new universality classes
of Anderson localization/delocalization transitions.  
Perhaps the most important is the Chalker-Coddington
network model for the quantum Hall transition (QHT)\cite{ChalkCodd},
which can be mapped onto disordered Dirac fermions\cite{LFSG,ChoChalker}.
A  partial list of other applications includes to 
dirty superconductors
\cite{NTW,SenFis,BocSerZ,AlSiZirn}, and  studies of   
hopping models on bipartite lattices \cite{HWK}.  
More recent applications are to graphene \cite{graphene1,graphene2},
where the Dirac fermions are present from the start.  
The possible universality classes of disordered Dirac fermions
were  classified according to their discrete symmetries in \cite{BLclass}.
The latter classification contains 13 classes and is thus  a minor 
refinement of Altland-Zirnbauer's classification which does not
assume the Dirac structure\cite{AltlandZirnbauer}.

A number of new theoretical techniques have been developed 
over the last decade to study these 
problems;  a partial list includes
for instance 
\cite{Mudry1,Bernard1,Serban,Guruswamy,SpinCharge,Tsvelik2,FendleyKonik}.  
For the most part, a proper understanding of the critical
points for generic disorder is still lacking.   
A notable exception is the spin quantum Hall transition (SQHT).  
Its network model\cite{Chalker} can also be mapped onto disordered Dirac
fermions\cite{Senthil1}.  
Remarkably, the equivalent spin chain was
mapped onto 2D classical percolation 
by Gruzberg, Ludwig and Read\cite{Gruzberg,CardySQHT,CardySQHT2}, 
and this leads to the exact  knowledge of the correlation
length exponent $\nu_{\rm perc} = 4/3$ and 
density of states exponent $\rho (E) \sim E^{1/7}$.

For the QHT, one  should also mention the replica sigma model approach of
Pruisken\cite{Pruisken}.  Although it appears  
to have the right ingredients as outlined  in \cite{Khmel},  it  has proved
too difficult to solve thus far,  so it remains unknown whether
it really does have the correct critical point.  There is also  
 the later  proposal of Zirnbauer\cite{ZirnQHE} which uses supersymmetry.       
   Based on symmetry
and various other requirements the critical point for
the QHT was proposed to be described  by a  
sigma model of 
WZNW type based on  the supergroup $PSL(2|2)$.  
The model was further studied in \cite{TsvelikQHE}.  
The main problem with this proposal is that the level $k$ of
the $PSL(2|2)$ WZNW model is an exactly marginal perturbation
so that the model actually has a line of fixed points depending on $k$.   
This would lead to the prediction of non-universality 
in the QHT,  which is contrary to the numerical evidence.
(For a recent review, see \cite{KramerReview,MirlinReview}.)  
It was pointed out recently by Tsvelik that the value $k=8$ gives
very reasonable exponents\cite{TsvelikPsl}.  
(The identical  exponents were actually already speculated   in 
\cite{LeClairSC,networkRG}.) 
Unfortunately,   it was noted that 
there are no known  constructive arguments leading to $k=8$ 
based on the $PSL(2|2)$ approach.       In the work we present here, 
 $PSL(2|2)$ will not play a  r\^ole,
but rather the simpler superalgebra  $gl(1|1)$ will be central,  and we will
describe  a precise mechanism for obtaining higher integer levels
$k$ based on a new form of
super spin-charge separation\cite{gl11}.

Due to an extensive effort over the last few decades,  
vast classes of conformally invariant $2D$ critical points
can be constructed\cite{CFT} and many of these mathematical
constructions can be extended to theories with supergroup
symmetries.   However in the study of disordered Dirac fermions,
it is important that the possible critical point is supposed
to be reached by renormalization group flow in the effective
disorder-averaged effective field theory.    There are
a very limited number of known mechanisms for obtaining
a fixed point from a renormalization group flow,  
and unless disordered systems depend on some new mechanism,
it is helpful to identify the known mechanisms as a guide:

\n (i)  Non-linear sigma models like the $O(N)$ model.  Due to the 
Mermin-Wagner theorem, these models are only critical for 
$-2 < N < 2$.   

\n (ii) Non-linear sigma models with topological terms,  
the primary example being the $O(3)$ non-linear sigma model
with $\theta = \pi$.   

\n  (iii)  Certain relevant perturbations that induce flows
between minimal models.  

\n  (iv)  Decoupling of massive degrees of freedom using
spin-charge separation in the case of marginal current-current
interactions.   Here the primary example is the $1d$ Hubbard model
which has $SU(2) \otimes SU(2)$ symmetry.  One $SU(2)$ sector
is marginally relevant,  the other marginally irrelevant. 
Thus one $SU(2)$ is gapped out in the flow,  and the fixed point
is the $SU(2)$ current algebra theory (WZNW model)\cite{Affleck}.

For the sake of comparison with our work,  Pruisken's model
is based on the scheme (ii),  whereas Zirnbauer's model
is mathematically constructed directly at the fixed point
so it is unknown under which renormalization group flow scheme
it can be realized.  
In the approach pursued in the present work,  disorder averaging
of Dirac fermions is known to yield marginal current-current interactions
so naturally the mechanism for obtaining a fixed point will be (iv),
i.e.  based
on super version of spin-charge separation.     

The starting point of the present work is the detailed 
disordered  Dirac fermion theories for the Chalker-Coddington
network model for the QHT and its variant for the SQHT.  
Based on extensive numerical evidence\cite{MirlinReview}, 
there is no doubt that these specific models have a critical point. 
Performing the disorder averaging using Efetov's supersymmetric
method\cite{Efetov} is known to lead to marginal anisotropic  left-right
current-current interactions of the underlying super current 
algebra.    The problem with such marginal
perturbations is that they typically do not have perturbative
fixed points at finite values of the coupling constants.  
In particular,  the coupled beta functions do not have
any non-trivial zeros at one loop.  Higher loop corrections
to the renormalization group (RG) beta functions were computed
for the network models in \cite{LeClairSC,networkRG} based
on the general proposal in \cite{Gerganov} and also did
not reveal any perturbative fixed points.  These analyses 
were nevertheless useful for understanding whether any
new couplings were generated under RG.   
In these studies it is significant  that the couplings
flow  to a singular point in a finite RG time,  which 
suggests an incomplete resolution of the flow rather than
the lack of a fixed point.   
It was pointed out
that the higher order beta functions are possibly not exact due to some 
contributions that were missed at 4-loops\cite{Ludwig4loop},   however
it seems unlikely that this could resolve the issue in
a constructive manner.

Since a perturbative fixed point of the beta functions is
unlikely,  one must identify the correct non-perturbative mechanism
that singles out the expected fixed point.   
In this paper we propose to resolve the RG flow in two stages. 
We first focus on the important symmetries of the N-copy  theory
{\it before disorder averaging},  i.e. we identify the relevant
symmetries that are present for {\it any}  realization of the disorder.  
This leads to the special r\^ole of the symmetries corresponding
to the current algebras  
$gl(1|1)_N \otimes su(N)_0$ for the QHT and 
$osp(2|2)_{-2N} \otimes sp(2N)_0$ for the SQHT.   
(Our nomenclature is that $\lie_k$ refers to the current
algebra for the finite (super)  Lie algebra $\lie$ at level $k$. )  
The disorder averaged effective actions have several couplings
 which correspond to the strengths (variances)  of the various random
potentials.  Rather than study the simultaneous flow of all couplings,
in the first stage we set some of the couplings to zero and carry
out the RG flow for a subset of the couplings corresponding to the
above symmetries.   A new form of the super spin-charge separation
obtained in \cite{gl11} is then used to argue that in 
the first stage one flows to the fixed point $gl(1|1)_N$ for the QHT.  
This result indicates that the locality constraints 
studied in \cite{Mudry1} for the $gl(1|1)_k$ theory,
which led to $k$ being an inverse integer, are too restrictive. 
For the N-copy SQHT,  the analogous flow is to 
to $osp(2|2)_{-2N}$.  This kind of flow for 1-copy of the 
SQHT was studied in \cite{SpinCharge} where it was viewed as a fine-tuning
of the model.  In this paper our  point of view is that
the first stage of the RG flow identifies the proper degrees
of freedom that are the most important for the actual critical point.  

In the second stage of the flow, we restore the additional
kinds of disorder that were possibly initially present in
the model as additional relevant perturbations.   
The possible  operators which appear in this second stage are dictated by
the quantum numbers of the original fields and the super spin
charge separation.    Another new aspect of the present work
is that we use the results in \cite{gl11} to explicitly
construct the operators corresponding to the additional
kinds of disorder.  In particular,  $gl(1|1)_k$ at any level $k$
has a simple free field representation in terms of 
two scalar fields and a symplectic fermion.    
The additional kinds of disorder correspond to logarithmic 
operators of scaling dimension zero.  For the QHT one obtains
a $gl(1|1)$ generalization of the sine-Gordon theory,
where $N$, the number of copies, appears as a coupling.
It can also be viewed as radius of compactification $R=\sqrt{N}$.    
The important feature of these kinds of perturbations is
that they do not drive the theory to a new fixed point, but
rather just lead to logarithmic corrections to the correlation
functions\cite{CauxLog,gl11}, and this explains why
for example the $osp(2|2)_{-2}$ current algebra contains
the correct exponents for the SQHT.

As models of disordered Dirac fermions,  the QHT and SQHT
are not so different in their formulations, and if
the methods are general enough, they should be subject to
the same kind of analysis.   It is therefore very instructive
to work out both cases in parallel, since some exact
results are known for the SQHT.   This also avoids 
idiosyncratic proposals for special cases.   
All the remaining sections of this paper have subsections
treating the QHT and SQHT cases.

Our results are presented as follows.   
 In sections II and III
we review the definitions of the models and the supersymmetric
method for disorder averaging, introducing a convenient 
notation to deal with the profusion of fields in the N-copy theories.
The symmetries of the models for any realization of disorder
are studied in section IV.  In section V we consider
a subgroup of these symmetries that commutes with the
permutation of the $N$ copies,  which   leads  to 
$gl(1|1)_N$ and $osp(2|2)_{-2N}$. 
The properties of these super current algebras that we need
are reviewed in section VI.      
In section VII we describe our 2-stage strategy for resolving
the RG flow.   Since we focus on symmetries that are present
for any realization of disorder,  the analysis does not
depend strongly on any assumed distributions of the random potentials.
 On the other hand the arguments rely strongly  on super spin-charge separation
and some simple 1-loop beta function arguments concerning the
marginal relevance/irrelevance of operators in the disorder
averaged effective action.   The additional perturbations
in the second stage of the RG flow are constrained by the
quantum numbers of the fields after gapping out the 
$su(N)_0$ and $sp(2N)_0$ ``copy'' symmetries.   Typically one
obtains perturbations by 
logarithmic operators which were explicitly constructed
in \cite{gl11}.   This results in some relatively simple
lagrangians involving the two scalar fields and a symplectic fermion.
Under some assumptions,  
the multi-fractal exponents are computed in section VIII.     Our 
results agree favorably (within  about $1\%$) with the numerical simulations 
in \cite{Klesse,MirlinQHT,MirlinSQHT,MirlinSQHT2}.  
In section IX we discuss the localization length exponents.

\section{Definition of the models.}

\subsection{Chalker-Coddington network model.}

The Chalker-Coddington network model can be mapped to the
following $2d$ hamiltonian\cite{ChoChalker,LFSG}
\beq
\label{2.1}
H = \( \matrix{  V + M & -i \d_\zbar + A_\zbar \cr
-i \d_z + A_z & V-M \cr } \) 
\eeq
where $z,\zbar$ are euclidean light-cone coordinates,
$z=(x+iy)/\sqrt{2}, \zbar = z^*$ with $x,y$ the $2d$ spacial
coordinates.   $A_{z,\zbar}$ is a $u(1)$ gauge field,
$A_z = (A_x + i A_y)/\sqrt{2}$.  The hamiltonian is hermitian
if $A_x, A_y, V$, and $M$ are real.    
The hamiltonian is first order in derivatives and operates
on a 2-component wave-function.   It thus corresponds to 
a universality class of disordered Dirac fermions,  
 class {\bf A=GUE} in \cite{AltlandZirnbauer},  or class {\bf 0} according
 to the more specific classification in \cite{BLclass}.     
All the potentials $A, M , V$ depend on $x,y$ and are random variables.  
The model in \cite{Guruswamy} on the other hand is in the 
chiral {\bf GUE},  i.e. class {\bf 2}.  
\subsection{Spin network model.}

The network model for the SQHT is also a model of diordered
Dirac fermions,  but in class
{\bf C}  \cite{AltlandZirnbauer} (class  {\bf $4_-$ } according to\cite{BLclass}).  
  The hamiltonian is
\beq
\label{2.2}
H = \( \matrix{ 2 \alphavec\cdot \sigmavec + M  & -i \d_\zbar +
A_\zbar \cr 
-i \d_z + A_z & 2 \alphavec\cdot \sigmavec - M \cr } \)
\eeq
where $\sigmavec$ are Pauli matrices and $A$ is an $su(2)$ gauge
field, $A = \Avec \cdot \sigmavec$.   
The hamiltonian thus operates on a 4-component wave-function.  
Again, all the potentials $\alphavec, M$, and $A$ depend on $x,y$
and are random.

\section{Supersymmetric Disorder averaging.}

Since the hamiltonians describe non-interacting fermions, the
disorder averaged correlation functions can be studied with
Efetov's supersymmetric method\cite{Efetov}.  

\subsection{QHT} 

Let us denote the 2-component wave-functions as follows:
\beq
\label{3.1}
\psi = \( \matrix{ \psi_+ \cr \psibar_+ \cr } \) , ~~~~~
\psi^\star = ( \psibar_- , \psi_- ) 
\eeq
For simplicity, let us refer to all  the disordered potentials
simply as ``$V$''. 
   The Green functions can be defined with respect to a functional integral
with the action
\beq
\label{3.1b}
S(\psi; V) = i \int \frac{d^2 x}{2\pi} ~ \psi^\star \, H (V)  \, \psi 
\eeq
Above, the fields are taken to be fermionic.  

The supersymmetric method is a trick to cancel the fermionic
determinant $Z(V)$ where $Z$ is the partition function at 
fixed disorder $V$.   One introduces ghost partners $\beta, \beta^\star$
 to
the $\psi$'s
and considers the action
\beq
\label{3.2}
S_{\rm susy}  = S(\psi ; V) + S(\beta; V)
\eeq
where $S(\beta; V)$ is identical to $S(\psi; V)$ but
with the replacement $\psi \to \beta$.   The $\beta$-fields
are bosonic.  
The effective action upon disorder averaging is then
defined as
\beq
\label{3.3}
\rme^{-S_{\rm eff} (\psi, \beta)} = \int DV\, \CP[V] \, \rme^{-S_{\rm susy}} 
\eeq
where $\CP[V]$ is the probability distribution of the random
potentials.    If $\CP[V]$ is taken to be gaussian, then 
$S_{\rm eff}$ contains quartic interactions among the fermions
and ghosts.    As we will see,  many of our arguments are independent
of the specific form of these probability distributions.

In order to clearly display the symmetries of $S_{\rm susy}$
and $S_{\rm eff}$, it will prove convenient to introduce the 
following notation.  Let $\Psi_\pm$ denote 2-component fields
built of out $\psi_\pm, \beta_\pm$,  and similarly for $\Psibar_\pm$:
\beq
\label{3.4}
\Psi_\pm = (\psi_\pm , \beta_\pm ), ~~~~~
\Psibar_\pm = (\psibar_\pm , \betabar_\pm )
\eeq
The index that runs over the two components of $\Psi_\pm$ will
be denoted as ``$r$'', $r=1,2$:
$\Psi^1_\pm = \psi_\pm, ~ \Psi^2_\pm = \beta_\pm$.  

We will   also be interested in computing disorder averaged moments
of correlation functions.  To compute averages of
$N$-th moments such as
\beq
\label{moments}
\bar{ \langle \psi (x) \psi (0) \rangle \,  \langle \psi (x) \psi (0) \rangle 
\ldots \langle \psi (x) \psi (0) \rangle }
\eeq
we need to introduce $N$-copies of the models.
Namely, we introduce fields $\Psi^\alpha_\pm$, $\alpha = 1,.., N$,
so that the complete set of fields is $\Psi_\pm^{r;\alpha}$
and $\Psibar_\pm^{r;\alpha}$, $r=1,2$. 
Thus  $\Psi_+$ refers
to $2N$ different fields.   

At  a fixed realization of disorder, $S_{\rm susy}$ can be
expressed in the compact form:
\barray
\nonumber  
S_{\rm susy} &=& \int \frac{d^2 x}{2\pi} [ ~  
\Psibar_- ( \d_z - i A_z (x)  ) \Psibar_+  + 
\Psi_- (\d_\zbar - i A_\zbar (x) ) \Psi_+ 
- i V(x)  \( \Psibar_- \Psi_+ + \Psi_- \Psibar_+ \) 
\\ \label{susyQHT}
&~&~~~~~~~~~~~~~~~~~~~~~~~~~~~~~~~~~~~~~- i M(x)  \( \Psibar_- \Psi_+ - \Psi_- \Psibar_+ \) 
 ~ ] 
\earray
where for example $\Psibar_- \Psi_+ = \sum_{r, \alpha} 
\Psibar_-^{r;\alpha} \Psi_+^{r;\alpha} $.

\subsection{SQHT}

For the spin-network model, one needs to introduce an 
additional $su(2)$ index ``$i$'' and consider fields
$\Psi_{\pm , i}^{r;\alpha}$, i.e. there are $4N$ fields
in $\Psi_+$ for example.   The action is then
\barray
\nonumber 
S_{\rm susy} &=&  \int \frac{d^2 x}{2\pi}  [~  
\Psibar_- ( \d_z - i A_z (x)  ) \Psibar_+  + 
\Psi_- (\d_\zbar - i A_\zbar (x) ) \Psi_+ 
- i  \alphavec (x) \cdot \( \Psibar_-  \sigmavec \Psi_+ + \Psi_- 
\sigmavec \Psibar_+ \) 
\\ \label{susySQHT}
&~& ~~~~~~~~~~~~~~~~~~~~~~~~~~~~~~~~~~~- i M(x)  \( \Psibar_- \Psi_+ - \Psi_- \Psibar_+ \) 
~ ] 
\earray
Above, the Pauli matrices, including the ones in $A$,  operate
on the index $i$  so that for example 
$\Psi_- \sigmavec \Psibar_+ = \sum_{r,i,j,\alpha} 
\Psi_{-,i}^{r;\alpha} \sigmavec_{ij} \Psibar_{+,j}^{r;\alpha} $.

\section{Symmetries at fixed disorder.}

\subsection{QHT}  

First consider all disordered potentials set to zero in $S_{\rm susy}$.  
The result is a free conformal field theory of Dirac fermions and
ghosts which has total central charge equal to zero:
\beq
\label{Sfree}
S_{\rm free} = \int \frac{d^2 x}{2\pi} 
\sum_{\alpha=1}^N \(  \psibar_-^\alpha  \d_z  \psibar_+^\alpha 
+ \psi_-^\alpha \d_\zbar \psi_+^\alpha   + 
\betabar_-^\alpha \d_z \betabar_+^\alpha 
+ \beta_-^\alpha  \d_\zbar \beta_+^\alpha \)
\eeq
   The two
point functions are
\beq
\label{4.1}
\langle \psi_- (z) \psi_+ (w) \rangle = 
\langle \psi_+ (z) \psi_- (w) \rangle =
\langle \beta_+ (z) \beta_- (w) \rangle =
- \langle \beta_- (z) \beta_+ (w) \rangle
= \inv{z-w}
\eeq
and similarly for the right-movers, 
$\langle \psibar_-(\zbar) \psibar_+ (\bar{w}) \rangle =
1/(\zbar - \bar{w} )$, etc. 
(For a review of $2D$ conformal field theory see \cite{CFT,Ginsparg}.)  
In the sequel we will not display the right-moving counterparts
if they are the obvious duplications of the left.  

For notational simplicity let us group the $r,\alpha$ indices
of $\Psi_\pm^{r;\alpha}$ into a single index $a$ and refer to
these fields as $\{ \Psi_\pm^a \}$, $a=1,..,2N$.    The extra
minus sign in the above two point functions can be accounted
for by introducing a grade $[a]=0$ for bosonic components 
and $[a]=1$ for fermionic ones.    One then has
\beq
\label{4.2} 
\Psi^a_+ (z) \Psi_-^b (0) \sim \inv{z} \, \delta^{ab} , ~~~~~
\Psi^a_- (z) \Psi_+^b (0) \sim  \inv{z} \,(-)^{[a]+1}  \delta^{ab}
\eeq
The complete set of chiral currents are then
\beq
\label{4.3}
J_\pm^{ab} = \Psi_\pm^a \Psi_\pm^b , ~~~~~~
H^{ab} = \Psi_+^a \Psi_-^b 
\eeq
These currents generate an $osp(2N|2N)_k$ 
super-current algebra at level $k=1$,  and this represents
the maximal symmetry without disorder.      Our
conventions for all the current algebras that appear in
this paper are presented in Appendix A.  

Without disorder the symmetry is actually the sum of left and right,
$osp(2N|2N)_1^L \oplus   osp(2N|2N)_1^R$ since the theory is conformal. 
In the presence of disorder  the conformal symmetry is broken
and one does not have the full current-algebra symmetry.
However one can study the global left-right diagonal symmetries
generated by the charges
\beq
\label{charges}
Q = \oint \frac{dz}{2\pi i} ~ J(z)  + \oint \frac{d\zbar}{2\pi i} ~
\bar{J} (\zbar) 
\eeq
where $J, \bar{J}$ are the  left/right moving currents.  
These conserved charges are always associated to a set of
left-moving currents for a current algebra 
and in the sequel this correspondence is implicit.  

With disorder,  the maximal 
$osp(2N|2N)$  symmetry is 
broken to something smaller.   Consider the transformation
$\psi \to \psi + \delta \psi$  which acts left-right diagonally:
$\delta \psi_-^\alpha = \beta_-^{\alpha'} , ~
\delta \beta_+^{\alpha'} = - \psi_+^\alpha$,  
 $\delta \psibar_-^\alpha = \betabar_-^{\alpha'} , ~~
\delta \betabar_+^{\alpha'} = - \psibar_+^\alpha$, and is
zero on all other fields, where $\alpha, \alpha'$ are fixed copy
indices.   
All of the operators $\Psi_- \Psi_+ , \Psibar_-  \Psibar_+ , \Psi_- \Psibar_+$
and $\Psibar_- \Psi_+$  are  invariant under this transformation.  
There is another symmetry of this type with $+ \leftrightarrow -$.  
The left-moving currents that generate these two symmetries
are
$S_\pm^{\alpha,\alpha'} = \pm \psi_\pm^\alpha \beta_\mp^{\alpha'}$. 
The charges for these fermionic symmetries 
 are nilpotent, 
$Q^2 = 0$,  and the symmetry they generate can thus
be thought of as a BRST symmetry.  Namely,  the 
disorder dependent part of $S_{\rm susy}$ can be written
as $\delta X$ for some $X$,  and its invariance is a consequence
of $\delta^2 = 0$\cite{Bernard1}.    

Consider  other nilpotent symmetries with 
$\delta \psi_-^\alpha = \beta_+^{\alpha'} $, i.e. that flip
the $u(1)$ charges.   These correspond to the currents 
$\Shat_\pm^{\alpha,\alpha'} = \psi_\mp^\alpha \beta_\mp^{\alpha'} $.  
One finds in this case that due to fermionic exchange signs,
the operators $\Psi_- \Psi_+$ and $\Psibar_- \Psibar_+$  
are not invariant.    The only invariant is the combination 
$(\Psibar_- \Psi_+ - \Psi_- \Psibar_+ )$.  

Thus, examining the action eq. (\ref{susyQHT}), one sees that
the only nilpotent symmetries are those corresponding to the diagonal
left-right symmetry which corresponds to the left-moving
currents $S_\pm^{\alpha, \alpha'} $.   The operator product expansion
(OPE) of these currents closes on the super-current algebra
$gl(N|N)_{k=1}$. (See Appendix A.)
   In other words,  at any fixed realization 
of the disorder,  the model has a global $gl(N|N)$ symmetry
corresponding to the current algebra $gl(N|N)_1$.   
We will refer to this  symmetry as the BRST symmetry.

\subsection{SQHT}

Since there are twice as many fields in the SQHT, the
maximal current algebra symmetry with zero disorder is
$osp(4N|4N)_1$.    Repeating the analysis above for the
QHT, one sees that there are nilpotent symmetries
generated by the currents  $S_\pm^{\alpha \alpha'} = 
\pm \sum_{i=1,2} \psi_{\pm, i}^\alpha \beta^{\alpha'}_{\mp , i} $.  

A  basic result  we will use repeatedly is the following.  
 Given two  copies of the same current 
algebra with currents $J^a_1$ at level $k_1$ and $J^a_2$
at level $k_2$ which furthermore commute, $[J^a_1 (z) , 
J^b_2, (w) ] = 0$.   Then $J^a = J^a_1 + J^a_2$ 
satisfies the current algebra at level $k_1 + k_2$.
Thus, since the $su(2)$ indices $i$ are summed over  in
$S_\pm$,  these currents close on $gl(N|N)_{k=2}$ since
each copy has level $1$ and the levels add.  

The $gl(N|N)$ symmetry is actually enlarged due to an additional
nilpotent symmetry.   Introduce the matrix $\ep$ which
acts on the $su(2)$ indices $i$:  
$\ep = \Big( \begin{array}{rc} 0~&1\\[-0.2cm] -1~&0\end{array}\Big)
$.
Using $\ep \sigmavec = - \sigmavec^t  \ep$ and $\ep^2 = -1$, one
can verify  that $S_{\rm susy}$ is invariant under the left-right
diagonal symmetry corresponding to the left-moving currents
$\Shat_\pm^{\alpha\alpha'} = \psi^\alpha_\pm \, \ep\,  \beta_\pm^{\alpha'}$. 
It is important to note that this symmetry would not be valid
if there were additional ``$V$'' type of disorder,  or if the
gauge field $A$ contained a $U(1)$ component.   
For $N=1$ copy,  the currents $\Shat_\pm,  S_\pm$ close
on the $osp(2|2)_k$ current algebra at level $k=-2$\cite{SpinCharge}.  
For $N$-copies this symmetry is promoted to the 
BRST symmetry $osp(2N|2N)_{-2}$.

\section{Permutation invariant BRST symmetries.}

The BRST symmetries discussed in the last section
are rather large since their dimension depends on
the number of copies $N$.     Furthermore,  the
current algebras are only moderately interesting
as possible critical points;  for example the 
$gl(N|N)_1$ theory has only integer scaling dimensions
at level $1$.       
In this section we constrain the possible fixed point further
by considering permutations in the number of copies.

Let $\CP_N$ denote the discrete permutation group for $N$ elements.
The actions $S_{\rm susy}$ possess this symmetry where $N$ is
the number of copies.    It is natural then to make
the hypothesis that a possible fixed point also has the
permutation symmetry.       The
BRST symmetries of the last section do not commute with $\CP_N$,
however there is a sub-algebra that does, which we will
refer to as the $\CP_N$ invariant BRST symmetry.   In section
VII  we will provide arguments based on super spin-charge
separation that indicate how  a fixed point with this restricted symmetry can
arise under RG flow.  

\subsection{QHT}

For the QHT  the generators that commute with $\CP_N$ are
\beq
\label{5.1}
H = \sum_\alpha  \psi_+^\alpha \psi_-^\alpha, ~~~~~
J = \sum_\alpha \beta_+^\alpha \beta_-^\alpha, ~~~~~
S_\pm = \pm \sum_\alpha \psi_\pm^\alpha \beta_\mp^\alpha
\eeq
The above currents satisfy the $gl(1|1)_k$ current algebra
at level $k=N$:
\barray
\nonumber
H(z) H(0) &\sim& \frac{k}{z^2} ,~~~~~ J(z) J(0) \sim - \frac{k}{z^2}
\\ 
\label{5.2}
H(z) S_\pm (0) &\sim& J(z) S_\pm (0) \sim \pm \inv{z} ~ S_\pm  
\\
\nonumber
S_+ (z) S_- (0) &\sim& \frac{k}{z^2} + \inv{z} ~ (H - J ) 
\earray

It will be important to determine any additional continuous  symmetries
that commute with the $\CP_N$-invariant BRST symmetry $gl(1|1)_N$.  
There is obviously an $su(N)$ symmetry which mixes the copies.  
Let $L^a_\psi, L^a_\beta$ denote the $su(N)$ currents
in the separate sectors and $L^a$ their sum:
\beq
\label{5.3}
L^a_\psi = \psi_-^\alpha t^a_{\alpha \alpha'} \psi_+^{\alpha'}, ~~~~~
L^a_\beta = \beta_-^\alpha t^a_{\alpha \alpha'} \beta_+^{\alpha'} , 
~~~~~L^a = L^a_\psi + L^a_\beta
\eeq
where $t^a$ is the $N\times N$ dimensional matrix representation 
of the vector  of $su(N)$.    The currents $L_\psi$ satisfy
$su(N)_1$, whereas the $L_\beta$ satisfy $su(N)_{-1}$. 
Therefore the  total currents  $L^a$ satisfy $su(N)_0$ at level $k=0$.  
In summary,  the symmetries that will play a significant r\^ole 
in the sequel is $gl(1|1)_N \oplus su(N)_0$ and these two 
current algebras commute.

\subsection{SQHT}

For the SQHT the $\CP_N$ invariant BRST symmetries correspond
to the currents
\barray
\label{5.4} 
H &=& \beta_+ \beta_- ,  ~~~~J = \psi_+ \psi_- , ~~~~~
J_\pm = \psi_\pm\,  \ep\, \psi_\pm 
\\ \nonumber
S_\pm &=&  \psi_\pm \beta_\mp , ~~~~~
\Shat_\pm = \psi_\pm \, \ep \, \beta_\pm 
\earray
where $\psi_+ \psi_- = \sum_{i,\alpha} \psi_{+,i}^\alpha \psi_{-,i}^\alpha$
and 
$ \psi_\pm \, \ep \,  \psi_\pm = \sum_{i,j,\alpha} \psi_{\pm, i}^\alpha \,
 \ep_{ij} \, 
\psi_{\pm, j}^\alpha$, etc.  
The above currents satisfy $osp(2|2)_{-2N}$.

The $\Shat_\pm$ and $J_\pm$ currents are invariant
under $\psi_\pm \to M \psi_\pm$, $\beta_\pm \to M \beta_\pm$ 
where $M$ is a $2N$ dimensional matrix satisfying
$M^t( \ep \otimes 1 ) M = \ep \otimes 1$.  $M$ is thus
an element of $Sp(2N)$.   The currents satisfy 
$sp(2N)_0$.   Since $\Shat_\pm$ and $J_\pm$ close
on $osp(2|2)_{-2N}$,   the $sp(2N)_0$ commutes with $osp(2|2)_{-2N}$

For $N=1$,  $sp(2) =su(2)$, and this $su(2)$ is 
the original $su(2)$ symmetry of the spin network model. 
We wish to emphasize that here the $sp(2N)_0$ symmetry 
is a property of the $N$-copy theory, which is to be contrasted with 
different   models that have a random $sp(2N)$ gauge 
field from the very beginning in the 1-copy theory, 
for example the models   in\cite{BernardSerban}.

\section{The $gl(1|1)_k$ and $osp(2|2)_k$ super current algebras.}

In this section we summarize the main results we will need
for the current algebras $gl(1|1)_k$ and $osp(2|2)_k$.
For $gl(1|1)_k$  we mainly summarize our recent  work\cite{gl11},
which builds on \cite{LudwigTwist,Schomerus1}.   
The $osp(2|2)$ results are based on the work\cite{Serban}.

\subsection{$gl(1|1)_k$}

We will need the Sugawara stress tensor $T(z)$.
The algebra $gl(1|1)$ has
two independent quadratic casimirs:
\beq
\label{casimirs}
C_2 = J^2 - H^2 + S^+ S^- - S^- S^+ , ~~~~~
C_2' = (J-H)^2 
\eeq
where it is implicit  that the above operators are the zero modes
of the currents.   The stress tensor is fixed by the condition
$T(z) J^a (0) \sim J^a (0) /z^2 $,  which requires it to be built
out of both casimirs\cite{RozSal}:
\beq
\label{glstress}
T(z)  = - \inv{2k} \( J^2 - H^2 + S_+ S_- - S_- S_+ \) 
+ \inv{2k^2} (J-H)^2 
\eeq

For any level $k$ there exists a free field representation
in terms of $2$ scalar fields and a symplectic fermion.
The free fields have the action
\beq
\label{6.1}
S = \inv{8\pi} \int d^2 x ~ \sum_{a,b=1}^2 \( 
\eta_{ab} \d_\mu \bfphi^a  \d_\mu \bfphi^b + 
\ep_{ab} \d_\mu \bfchi^a \d_\mu \bfchi^b  \) 
\eeq
where
\beq
\label{6.2}
\eta =  \(\matrix{1&0\cr 0 & -1 \cr}\), ~~~~~~
\ep = \( \matrix{0&1\cr -1& 0\cr} \)
\eeq
and $\d_\mu^2 = 2  \d_z \d_\zbar$.  The $\bfchi$ fields are
Grassman: $(\bfchi^a )^2 = 0$ and have Virasoro central charge $c=-2$,
so that the total central charge is zero.  
Note that the metric for the bosonic fields  has indefinite signature,
which will turn out to be important.  
The equations of motion imply that the fields can be decomposed
into left and right moving parts:
\barray
\label{6.3}
\bfphi^a (z ,\zbar ) &=& \phi^a (z) + \phibar^a (\zbar) 
\\ \nonumber
\bfchi^a (z, \zbar) &=& \chi^a (z) + \chibar^a (\zbar) 
\earray
Above,  the bold face signifies local fields.   The two point functions are
\beq
\label{6.4}
\langle \phi^a (z) \phi^b (w) \rangle = - \eta^{ab} \log (z-w), 
~~~~~
\langle \chi^a (z) \chi^b (w) \rangle = - \ep^{ab} \log(z-w) 
\eeq
(Our conventions are $\eta^{ab} = \eta_{ab}, \ep^{ab} = \ep_{ab}$.)
Exponentials of the bosons have the conformal dimension:
\beq
\label{vert}
\Delta \( \rme^{i (a \phi^1 + b\phi^2 )} \) = \frac{a^2 - b^2}{2}
\eeq

It is straightforward to verify the following representation of
the OPE's in eq. (\ref{5.2}):
\barray
\label{6.5}
H &=& i \sqrtk \, \d_z \phi^1 , ~~~~~ J = i \sqrtk \, \d_z \phi^2 
\\ \nonumber
S_+ &=& \sqrtk \, \d_z \chi^1\,  \rme^{i
 (\phi^1 - \phi^2 )/\sqrtk }
,
~~~~~
S_- = - \sqrtk \, \d_z \chi^2 \, \rme^{ -i
 (\phi^1 - \phi^2 )/\sqrtk  }
\earray
In the sequel, where there is no cause for confusion, we
will simply write $ \d \phi$ for $\d_z \phi (z)$.

The algebra $gl(1|1)$ has 2-dimensional representations
where $H = {\rm diag}(h,h-1)$ 
and $J = {\rm diag}(j,j-1)$ which will
be denoted as $\glrep{h}{j}$.  (We follow the conventions in \cite{gl11}.)
These are so-called typical representations when $h\neq j$.    
Primary fields associated with these representations have
conformal dimension
\beq
\label{6.5b}
\Deltahj{h}{j} = \frac{ (h-j)^2 }{2k^2} + \frac{ (h-j) (h+j -1)}{2k} 
\eeq
The basic  fields $\psi_\pm, \beta_\pm$ 
are in the fundamental representations:
\beq
\label{6.5c}
(\psi_+ , \beta_+ ) \leftrightarrow \glrep{1}{0} , ~~~~~
(\beta_- , \psi_- ) \leftrightarrow \glrep{0}{1}
\eeq
and have scaling dimension $1/2$ when $k=1$. 

The tensor product of two typical representations 
is 
\beq
\label{6.6}
\glrep{h_1}{j_1} \otimes \glrep{h_2}{j_2} = 
\glrep{h_1 + h_2}{j_1 + j_2 } \oplus
\glrep{h_1 + h_2 -1}{j_1 + j_2 -1} , ~~~~~
(h_1 + h_2 \neq j_1 + j_2 )
\eeq
When $h_1 + h_2 = j_1 + j_2$ the tensor product gives a new
reducible but indecomposable 4-dimensional  representation denoted $\hfour{h}$:
\beq
\label{6.7}
\glrep{h_1}{j_1} \otimes \glrep{h_2}{j_2} = 
\hfour{h_1 + h_2 -1}
\eeq
These representations have $\Delta_{\hfour{h}} = 0$, however they
are logarithmic since the casimir $C_2$ is not diagonal.  

The vertex operators $\V{h}{j}$ can be explicitly constructed
in the free field theory,  and require  the twist
fields in the symplectic fermion sector\cite{gl11}.   
As for the spin fields of the Ising model,  the 
twist fields modify the boundary
conditions of the fundamental field $\chi$:
\barray
\label{6.8}
\chi^1 ( \rme^{2\pi i } z )  \mu_\lambda (0) &=& \rme^{-2\pi i \lambda} \chi^1 (z) \mu_\lambda (0) 
\\ \nonumber
\chi^2 ( \rme^{2\pi i } z )  \mu_\lambda (0) &=& 
\rme^{2\pi i \lambda} \chi^2 (z) \mu_\lambda (0) 
\earray
The properties of these fields were studied in\cite{Kausch}.  
It is clear from the above equation that $2 \pi \lambda$ is a phase and is restricted to
$-1 < \lambda < 1 $.  We also need the doublet of twist fields $\sigma_\lambda^a$,
 which arise in the OPE:
\beq
\label{sigs}
\d \chi^1 (z) \mu_\lambda (0) \sim \frac{\sqrt{1-\lambda}}{z^\lambda} \, 
\sigma^1_\lambda , ~~~~~
\d \chi^2 (z) \mu_\lambda (0) \sim \frac{\sqrt{\lambda}}{z^{1-\lambda}}  
\sigma^2_\lambda
\eeq
The conformal dimensions of the twist fields  are 
\beq
\label{dimtwist}
\Delta(\mu_\lambda) = \frac{\lambda(\lambda-1)}{2} \equiv \Deltachi_\lambda , 
~~~~~\Delta(\sigma_\lambda^1 ) =\Deltachi_{\lambda -1} , ~~~~~
\Delta(\sigma_\lambda^2 ) = \Deltachi_{\lambda +1}
\eeq
The
vertex operator for $\V{h}{j}$ requires twist fields with
$\lambda = (h-j)/k$.  For $h>j$:
\beq
\label{exvert1}
\V{h}j  = (h-j)^{1/4} \( \matrix{
-   \mu_\lambda ~  \rme^{i(h\phi^1 - j\phi^2)/\sqrtk}  \cr
\sigma^2_\lambda ~ \rme^{i( (h-1) \phi^1 - (j-1) \phi^2 )/\sqrtk } \cr }
\) , ~~~~~ \lambda = \frac{h-j}{k}
\eeq
whereas for $h<j$:
\beq
\label{exvert2}
\V{h}{j} = (j-h)^{1/4} \( \matrix{
\sigma_{1+\lambda}^1 ~  \rme^{i(h\phi^1 - j\phi^2)/\sqrtk} \cr
\mu_{\lambda+1} ~ \rme^{i( (h-1) \phi^1 - (j-1) \phi^2 )/\sqrtk } \cr }  \) 
, ~~~~~ \lambda = \frac{h-j}{k}
\eeq

The vertex operator $V_{\hfour{h}}$ for the representation $\hfour{h}$ 
is constructed from the logarithmic field 
$\ep_{ab} \chi^a \chi^b$:
\beq
\label{6.9}
V_{\hfour{h}} = \( \matrix{ 
\chi^1 ~ \rme^{i(h+1)(\phi^1 - \phi^2 )/\sqrtk } \cr
\sqrtk ~ \rme^{i h (\phi^1 - \phi^2)/\sqrtk } \cr
\inv{\sqrtk} \chi^1 \chi^2 ~   \rme^{ih(\phi^1 - \phi^2)/\sqrtk} \cr
\chi^2 ~ \rme^{i(h-1)(\phi^1 - \phi^2)/\sqrtk}  \cr } \)
\eeq
The two middle fields  form a logarithmic
pair  with $\Delta=0$.  It is important  that
the above logarithmic field has a simple and explicit construction  in
the second-order symplectic fermion theory;  this is not transparent
in the minimal model description of $c=-2$, nor in the first-order
description.  For a review of logarithmic conformal field theory,
see \cite{Gaberdiel,Flohr}.

A closed operator algebra is obtained when $k$ is an integer
and the spectrum of fields $\V{h}{j}$ is restricted to
$h,j$ integers satisfying 
\beq
\label{6.10}
-k \leq h-j \leq k
\eeq
This operator algebra can be viewed as generated by OPE's of
the fundamental vertex operators $\V{1}{0}$ and $\V{0}{1}$.    For instance:
\beq
\label{6.11o}
\V{1}{0} (z) \V{0}{1} (0) \sim \inv{z^{1/k^2}} \,  V_{\hfour{0}}  
\eeq

\subsection{$osp(2|2)_k$.}

The finite dimensional representations of $osp(2|2)$ can be labeled by
the $su(2)$ with generators $J, J_\pm $ and by the $u(1)$ charge $H$.   
The typical, irreducible representations will be denoted as 
$\osprep{b}{s}$ where $s\in \{ 0, \inv{2} , 1, 
\frac{3}{2},....\}$ is an $su(2)$ spin and $b=H/2$.  
These representations are $8s$ dimensional.   In order to describe their
$su(2) \otimes u(1)$ decomposition,  let $\surep{b}{s}$ denote the
$2s+1$ dimensional representation with $J/2  = s_3 = -s, -s+1, ..., s$
and $H=2b$.    The generic decomposition is 
\beq
\label{6.11}
\osprep{b}{s} = \surep{b}{s} \oplus \surep{b+\texthalf}{s-\texthalf}  
\oplus  \surep{b-\texthalf}{s-\texthalf} \oplus \surep{b}{s-1} 
\eeq

As for $gl(1|1)$, there are atypical, indecomposable but reducible
representations at $b^2 = s^2$.   The simplest is $8$ dimensional 
and arises in the following tensor product
\beq
\label{6.12}
\osprep{0}{\texthalf} \otimes \osprep{0}{\texthalf} = 
\osprep{0}{1} \oplus \eight 
\eeq
The $\eight$ has the quantum numbers of $\osprep{\inv{2}}{\inv{2}} \oplus
\osprep{-\inv{2}}{\inv{2}}$.

The stress tensor is built from the single quadratic
casimir:
\beq
\label{ospstress}
T_{osp(2|2)} = \inv{ 2(2-k) } 
\[ 
J^2 - H^2 - \inv{2} (J_+ J_- + J_- J_+ ) 
+ (S_+ S_- - S_- S_+ ) + (\Shat_- \Shat_+ - \Shat_+ \Shat_- ) 
\]
\eeq
and the typical representations with $b^2 \neq s^2$ have conformal
dimension
\beq
\label{6.13}
\Delta^{osp}_{[b,s]} =  \frac{ 2 (s^2 - b^2 )}{2-k} 
\eeq

At $k=-2$ there is a free field representation with the same
field content as in eq. (\ref{6.1})\cite{LudwigTwist}.  
This can be derived from the $gl(1|1)_2$ embedding\cite{gl11}. 
In fact:
\beq
\label{woo}
T_{osp(2|2)_{-2}} = T_{gl(1|1)_2} = T_{gl(1|1)_{-2}}
\eeq

\section{Disordered critical points: super spin-charge separation
and the renormalization group.}

\subsection{General strategy.}

If the random potentials are taken to be gaussian distributed,
e.g. $\CP[V] = \exp ( - \inv{4\pi  g_V} \int d^2 x ~ V(x)^2 )$, 
then the functional integrals over all the random potentials can
be performed and one obtains the general form
\beq
\label{7.1}
S_{\rm eff} = S_{\rm free} + \int  \frac{d^2 x}{2\pi} \sum_A g_A
\CO^A (x) 
\eeq
where $S_{\rm free}$ is the free action of the $\psi_\pm ,
\beta_\pm$ fields  eq.  (\ref{Sfree}), 
$g_A$ are variances that measure the strength of
the disorder, and $\CO^A$ are marginal operators.  
The operators $\CO^A$ can always be expressed as left-right
current-current perturbations,  i.e. they are of the form
\beq
\label{7.2}
\CO^A = \sum_{a,b} d^A_{ab} J^a \bar{J}^b 
\eeq
for some bilinears $d^A_{ab}$, and the currents $J^a$ are those
of the maximal symmetry,  $osp(2N|2N)_1$ for the QHT and
$osp(4N|4N)_1$ for the SQHT, and are bilinears in the 
fields $\psi_\pm , \beta_\pm$.      
Since the models have the BRST symmetries for any realization
of disorder,  the operators $\CO^A$ must be BRST invariant,
i.e. $S_{\rm eff}$ has a $gl(N|N)$ symmetry in the QHT and
an $osp(2N|2N)$ symmetry for the SQHT.   The perturbations
$\CO^A$ can thus be viewed as anisotropic interactions 
of the maximal current algebra that are BRST invariant. 

As discussed  in the introduction,  the perturbative RG for the
simultaneous flow of all the couplings $g_A$ does not reveal
a fixed point.   In order to resolve this difficulty,  we propose
to perform the RG in two stages,  with special attention paid
to the symmetries that exist at any realization of disorder. 
   
We will need the following general property.  
Consider two commuting current algebras
$\CG_A$ and $\CG_B$ with currents $J_{A},  J_{B}$.
Furthermore, let us suppose that the stress tensor for a given
conformal theory separates  as follows:
\beq
\label{7.3}
T_{\rm cft}  = T_{\CG_A}  + T_{\CG_B}
\eeq
Consider the perturbation of the conformal field theory
by left-right current-current perturbations:
\beq
\label{curr.1}
S = S_{\rm cft}  +  \int \frac{d^2 x}{2\pi}  \(  
g_A  \, J_{A} \cdot \bar{J}_{A} +  
g_B  \, J_{B} \cdot \bar{J}_{B}
\)
\eeq
where $J\cdot \bar{J}$ is the invariant built on the
quadratic casimir.   Since the currents commute,  the
RG beta-functions decouple;  to 1-loop the result is: 
\beq
\frac{d g_A }{d \ell} = C^{\rm adj}_A \, g_A^2 , ~~~~~
\frac{d g_B}{d \ell} = C^{\rm adj}_B \, g_B^2
\eeq
where $\ell$ is the logarithm of the length scale
and $C^{\rm adj}_A $ is the casimir for the adjoint representation 
of the finite dimensional part 
of $\CG_A$.   Let us suppose that the physical
regime corresponds to positive $g_{A,B}$.  
If  $C^{\rm adj}_B$ is positive, then the  coupling
$g_B$ is marginally relevant and the flow is to infinity. 
This is a massive sector and the $\CG_B$ degrees of freedom
are ``gapped out'' in the RG flow.  
If $C^{\rm adj}_A$ is negative,  then the  coupling $g_A$ is 
 marginally irrelevant.  This results in the  fixed point
characterized by the 
$\CG_A$ current algebra.   If the original conformal field theory 
corresponds to the current algebra $\CG_{\rm max}$,  then the 
fixed point  may be viewed as  the coset
$\CG_{\rm max}/\CG_B$.  This scenario was proposed for 
generic fixed points of marginal current-current perturbations in
\cite{LeClairSC},  however what was missing in the argument
was the spin-charge separation. The resulting coset
 is a somewhat trivial example of
the GKO construction\cite{coset} because of the separation  
of the stress tensor.

In the first stage of the RG flow  we consider only the disorder
couplings 
for the $\CP_N$ invariant BRST symmetries and the
additional ``copy'' symmetries.   The essential ingredient
here is the spin-charge separation\cite{SpinCharge,BernardSerban,gl11}.  
In the first stage we will obtain a critical point corresponding
to the $\CP_N$ invariant BRST symmetry.   In other words, in
the first stage of the RG flow we identify the  massless
degrees of freedom that are most important at the critical point. 
    In the second stage
of the RG flow,  we reintroduce the other kinds of disorder
as further relevant perturbations of these massless degrees of freedom.

\subsection{QHT}

The $\CP_N$ invariant BRST symmetry is $gl(1|1)_N$ and
the copy symmetry that commutes with it is $su(N)_0$.  
Remarkably there exists the following spin-charge separation\cite{gl11}:
\beq
\label{7.4}
T^\Ncopy_\free = -\inv{2} \sum_{\alpha=1}^N 
( \psi_-^\alpha \d_z \psi_+^\alpha + 
\beta_-^\alpha \d_z \beta_+^\alpha  ) =  T_{gl(1|1)_{k=N}} + T_{su(N)_0} 
\eeq
where the stress tensors $T_{\glhat_{N}}$ and $T_{su(N)_0}$ are
the Sugawara stress tensors for interacting theories.   
Simple checks of the above result are as follows.
First, all of the stress tensors have $c=0$.    
The free theory contains $4N$ fields $\psi_\pm^\alpha, \beta_\pm^\alpha$.
Under the $gl(1|1)\otimes su(N)$ they transform as
$(\glrep{1}{0} \oplus \glrep{0}{1} ) \otimes [{\rm vec}]$ 
where $[{\rm vec}]$ is the $N$-dimensional vector representation of
$su(N)$.   The later has conformal dimension 
$\Delta_{su(N)_k} = \frac{N^2-1}{2N(k + N)}$ at general level $k$,
whereas $\Deltahj{1}{0} = \Deltahj{0}{1} = \inv{2k^2}$.  
One sees that the dimensions add up properly:
$\Deltahj{1}{0} + \Delta_{su(N)_0} = 1/2 $,  as is appropriate
for the free $\psi_\pm , \beta_\pm$ fields.  

In the first stage of the RG flow we consider the action
of the form eq. (\ref{curr.1}) where $\CG_A = gl(1|1)_N$
and $\CG_B = su(N)_0$.   For $su(N)$,  $C^{\rm adj} >0$,
and it is gapped out in the flow.  
For $gl(1|1)_N$ current perturbations 
the situation is somewhat more subtle because
there are two quadratic casimirs\cite{Guruswamy}.  Consider
\beq
\label{fac2}
S = S_{\rm free}  + \int \frac{d^2 x}{2\pi}  \[
g  \, \( J\bar{J} - H\bar{H} + S_+ \bar{S}_- - S_- \bar{S}_+ \) 
+ g'  (J-H)(\bar{J} - \bar{H} ) \]  
\eeq
Then the 1-loop beta function for $g$ is zero, 
whereas $d g'  / d \ell = - g^2 $\cite{Guruswamy}.
(Here we  fixed the sign by setting some couplings to zero in
the more general result in \cite{LeClairSC}.  )  
    Therefore these 
$gl(1|1)$ current interactions are irrelevant.  
It is important that this is in contrast to  
 the situation for the 
model in the Gade-Wegner universality class 
which has the same beta function up to a sign\cite{Guruswamy},
$d g'/ d\ell = + g^2$, which implies the disorder in that
case is marginally relevant.  
This difference in sign is a consequence of the detailed form
of the hamiltonian in \cite{Guruswamy} 
which acts on a 4-component wavefunction 
rather than two.   The sign of the beta function can also
be flipped by considering an imaginary gauge potential.  
The higher loop corrections computed in \cite{LeClairSC} 
do not alter this picture.  
Thus in the first stage of the RG flow, one flows to the
fixed point $gl(1|1)_N$.

It is important to point out that 
one  feature of this scheme for ariving at the fixed point
$gl(1|1)_N$ is that the critical exponents now  depend on
$N$,  whereas in the original $gl(N|N)$ invariant theory 
they were independent of $N$.     This is due to the fact
that we have gapped out the $SU(N)_0$ sub-algebra of 
the $gl(N|N)_1$.  On the other hand,  the common assumption
in the literature is that the critical exponents should
be independent of $N$,  even though an RG scenario that
achieves this has not yet been proposed.  We will return
to this issue where it must be faced in section VIII on
the multi-fractal exponents.

Additional kinds of disorder can now be incorporated as follows.
The original currents for the maximal symmetry $osp(2N|2N)_1$,
eq. (\ref{4.3})
can be classified according to the $gl(1|1)_N \otimes su(N)_0$. 
Since the RG flow in the first stage gaps out the $su(N)_0$,
  what remains are the $gl(1|1)_N$ degrees
of freedom.   The possible $gl(1|1)_N$ representations the residual
currents fall into follows from the fact that they are
bilinears in the fields $\psi_\pm , \beta_\pm$, and the
latter correspond to the $\glrep{1}{0}$ and $\glrep{0}{1}$
representations, eq. (\ref{6.5c}).   Using
the tensor products
\barray
\nonumber
\glrep{1}{0} \otimes \glrep{1}{0} &=& \glrep{2}{0} \oplus \glrep{1}{-1}
\\ \label{7.5}
\glrep{0}{1} \otimes \glrep{0}{1} &=& \glrep{0}{2} \oplus \glrep{-1}{1}
\\ \nonumber 
\glrep{1}{0} \otimes \glrep{0}{1} &=& \hfour{0}
\earray
one sees that the $J^{ab}_\pm$ currents transform in
the $\glrep{2}{0}, \glrep{1}{-1}, \glrep{0}{2}$ and
$\glrep{-1}{1}$ representations,  whereas $H^{ab}$ 
transform in the $\hfour{0}$.  Note that  $\hfour{0}$ has
the same quantum numbers as the adjoint of $gl(1|1)$,
and represents what is left of the $gl(N|N)$ after 
gapping out the $su(N)_0$.    The conformal
dimensions of these  representations are
\barray
\nonumber
\Deltahj{2}{0} &=& \Deltahj{-1}{1} = \frac{2+N}{N^2}
\\ \label{7.6}
\Deltahj{0}{2} &=& \Deltahj{1}{-1} = \frac{2-N}{N^2} 
\earray
and $\Delta_{\hfour{0}} = 0$.

Let $\bfPhi_\rep (z,  \zbar) $ denote the $gl(1|1)$ invariant
local field associated with the representation $\rep$ of
$gl(1|1)$ with scaling dimension $2\Delta_\rep$.   It 
can be expressed as a product of left-right vertex operators
$V_\rep \cdot \bar{V}_{\bar{\rep}}$.    For general $N$,  the 
fields $\bfPhi_{\glrep{2}{0}} , \bfPhi_{\glrep{-1}{1}}, 
\bfPhi_{\glrep{0}{2}}$ and $\bfPhi_{\glrep{1}{-1}}$
are expressed in terms of the twist fields $\mu_\lambda$
and $\sigma^a_\lambda$ with $\lambda = 2/N , 1-2/N$.
Explicit expressions can be found in \cite{gl11}.   The
field $\bfPhi_{\hfour{0}}$ on the other hand requires only
the symplectic fermion and bosons,  and for $N\leq 2$ it is the most
relevant operator.    
If we keep only the most relevant operator, then 
we should consider 
  \barray
\label{glSG}
S &=& S_{gl(1|1)_N}  + g \int \frac{d^2 x}{8\pi}  ~ \bfPhi_{\hfour{0}} 
\\ \nonumber 
&=&   \int \frac{d^2 x}{8\pi} \(  
 \sum_{a,b=1}^2   \eta_{ab} \,  \d_\mu \phi^a \d_\mu \phi^b  +
\ep_{ab} \, \d_\mu \chi^a \d_\mu \chi^b   
~+ g \, 
 \chi^1 \chi^2 \,  
\cos \( (\phi^1 - \phi^2 )/ \sqrt{N}  \)  \)
\earray
where here, and henceforth, 
 $\phi, \chi$ are the local fields $\bfphi, \bfchi$.

The case of $N=2$ is distinct since both $\bfPhi_{\glrep{0}{2}}$
and $\bfPhi_{\glrep{1}{-1}}$ have $\Delta=0$ which is degenerate with 
the dimension of 
$\bfPhi_{\hfour{0}}$, so that the latter 
 is no longer the most relevant operator.
The explicit forms at $N=2$ are \cite{gl11}:
\beq
\label{zerotwo}
\bfPhi_{\glrep{1}{-1}} - \tilde{\bfPhi}_{\glrep{0}{2}} =
4 \chi^1 \chi^2 \, \cos \( (\phi^1 + \phi^2 )/\sqrt{2} \) 
+ 4 (\d_\mu \chi^1 \d_\mu \chi^2 ) (\chi^1 \chi^2 ) \,
\cos (\sqrt{2} \phi^2 ) 
\eeq
($\tilde{\bfPhi}_{\glrep{0}{2}}$  only differs from
$\bfPhi_{\glrep{0}{2}}$ by some fermionic exchange signs.)
For $N>2$ the operators $\bfPhi_{\glrep{0}{2}}$ 
become more relevant than $\bfPhi_{\hfour{0}}$ 
and may need to be included as additional perturbations in
eq. (\ref{glSG}).

The important feature of logarithmic perturbations such as 
eq. (\ref{glSG}) is that we believe they should not  drive the theory
to another fixed point, but rather just give logarithmic
corrections to correlation functions.  General arguments were
given in \cite{CauxLog} for {\it marginal}  logarithmic perturbations.
Here the logarithmic perturbation has dimension zero and is thus
strongly relevant so the arguments in \cite{CauxLog} do not necessarily
apply.  However arguments based on the fact that the perturbation
has exactly dimension zero were given in \cite{gl11} supporting 
the idea that one is not driven to a new fixed point.  
For the concrete model 
eq. (\ref{glSG}) this is easy to see since, due to the 
indefinite signature of the scalar fields,  the OPE of
the exponentials is regular:
\beq
\label{regular}
\rme^{i a (\phi^1 - \phi^2 )(z)} \, \rme^{i b (\phi^1 - \phi^2)(0)} 
\sim {\rm regular}
\eeq
This implies that in conformal perturbation theory 
the perturbation behaves like a mass term $\chi^1 \chi^2$,
and does not lead to contributions to the beta function at any
order in perturbation theory.   
A contribution to the beta function for $g$ would
require a singular term in the OPE of the form $\bfPhi_{\hfour{0}} (x) 
\bfPhi_{\hfour{0}} (0) \sim \bfPhi_{\hfour{0}}$,
however there is no such term.  
Although we cannot give any stronger arguments at this stage, 
in the sequel we will adopt the working hypothesis       
that dimension zero logarithmic perturbations essentially do not
change the critical exponents.

\subsection{SQHT}

The version of super spin-charge separation we need was proven
in \cite{BernardSerban}:
\beq
\label{SQHEsc}
T_{\rm free}^{2N-{\rm copy}} =  T_{osp(2|2)_{-2N}} + T_{sp(2N)_0}
\eeq
Here the check of the scaling dimensions goes as follows. 
The $8N$ fields $\psi_{\pm,i}^\alpha, \beta_{\pm, i}^\alpha$ 
transform as $[0,\inv{2}]^{osp} \otimes [{\rm vec}]$ under
the $osp(2|2)_{-2N} \otimes sp(2N)_0$ where the vector
representation of $sp(2N)$ is $2N$ dimensional.   
For general $k$ the latter has dimension
$\Delta_{sp(2N)_k} = \frac{2N+1}{4(k + N+1)}$.  
At level $k=-2N$ from eq. (\ref{6.13}) one has
$\Delta^{osp}_{[0,\inv{2}]}  = \inv{4(N+1)}$.  
Again one has $\Delta^{osp}_{[0,\inv{2}]} + \Delta_{sp(N)_0} = 1/2$,
as required.

Repeating the same arguments as for the QHT,  since $C^{\rm adj}_{sp(2N)} >0$
and $C^{\rm adj}_{osp(2|2)} <0$,   the first stage of
the RG flows takes us to the current algebra $osp(2|2)_{-2N}$. 
For $N=1$ this  RG flow was studied in greater detail in 
\cite{SpinCharge,networkRG},  where it was viewed as 
a fine-tuning of the initial model.  (The couplings
$g_c$ and $g_s$ in \cite{networkRG} correspond to the couplings
for the $osp(2|2)$ and $su(2)$ currents respectively.)
    
As before, in the second stage we restore other kinds of
disorder by examining the quantum numbers of the remaining
degrees of freedom.  
The residual currents are bilinears of fields in the $\osprep{0}{\inv{2}}$ 
thus their quantum numbers follow from eq. (\ref{6.12}).  
Since $\Delta^{osp}_{[0,1]} = \inv{N+1}$
and the $\eight$ has $\Delta = 0$,  for any $N$
the dimension zero logarithmic field $\bfPhi_{[8]}$ is the
most relevant operator.    Therefore, the additional kinds
of disorder should correspond to the logarithmic perturbation:
\beq
\label{7.9}
S = S_{osp(2|2)_{-2N}} + g_8  \int \frac{d^2 x}{2\pi} ~ \bfPhi_{[8]}
\eeq
where  $S_{osp(2|2)_{-2N}}$  formally represents the conformal field theory 
with the current algebra symmetry.  
For $N=1$ the detailed study of the effective action 
indeed leads to the same conclusion\cite{SpinCharge,networkRG}.  

For the special case of $N=1$,   $osp(2|2)_{-2}$ has a free field 
representation with the same content as for $gl(1|1)_k$, so
that $S_{osp(2|2)_{-2}}$ has the free field form 
in eq. (\ref{6.1})\cite{LudwigTwist,gl11}.  
The explicit form of the perturbation was given in \cite{gl11}:
\beq 
\label{10.10}
\bfPhi_{[8]} =  4\,  \chi^1 \chi^2 \,  \(  
\cosh(  ( \phi^1 - \phi^2 )/\sqrttwo )
 + \cosh(( \phi^1 + \phi^2 )/\sqrttwo )  \) 
+ 4  (\d_\mu \chi^1 \d_\mu \chi^2 ) ( \chi^1 \chi^2 ) ~ 
\cosh(\sqrttwo \phi^1) 
\eeq
For general $N$ the free field representation requires more
fields\cite{Quella}.

Since the additional forms of disorder correspond to the
above logarithmic perturbation,  as we argued above,
the exponents for the SQHT should be contained in the 
$osp(2|2)_{-2N}$ theory.     For the 1-copy theory,
this is more transparent using the $gl(1|1)_2$ embedding
since $T_{gl(1|1)_2} = T_{osp(2|2)_{-2}}$\cite{gl11}.  
The vector representation of $osp(2|2)$ corresponds
to the  $gl(1|1)_2$ fields 
$\bfPhi_{\glrep{1}{0}}$ and $\bfPhi_{\glrep{0}{1}}$ 
with scaling dimension $2 \Delta_{\glrep{1}{0}} = 1/4$, and 
this   determines the density of states exponent
$\rho (E) \sim E^{1/7}$, since $1/7 = \Delta/(1-\Delta)$
with $\Delta = 1/8$.   
The remaining low dimension fields are $\bfPhi_{\glrep{2}{1}}$ 
and $\bfPhi_{\glrep{1}{2}}$ with $\Delta = 5/8, -3/8$ respectively.  
The other fields have
dimensions  which differ by an integer from the fields
considered thus far.   The $\Delta=5/8$ field 
determines the correlation length exponent for
percolation: $\nu_{\rm perc} = (2(1-5/8))^{-1} = 4/3$. 
(In the $osp(2|2)_{-2}$ description, the $\Delta = 5/8$ field
is a descendant of the field $[\pm 1, \inv{2}]^{osp}$ 
with $\Delta = -3/8$.)
Both of these exponents agree with the exact results 
in \cite{Gruzberg}.   
Note that the $c=0$ minimal model field with $\Delta = 1/3$,
which determines the localization length for self-avoiding
walks, is not contained in the spectrum, which is consistent
with observations made in \cite{CardySQHT,ReadSaleur}. 
Our proposal appears to be consistent with other observations
made in \cite{ReadSaleur}, 
since, because of the logarithmic perturbation, 
the critical point is not strictly speaking
a conformal current algebra, even though it has
some of the same exponents. 
A further check will be given in the next section
based on the multi-fractal exponents.  

On the other hand,  the above spectrum does not contain
the full spectrum for the SQHT proposed in\cite{ReadSaleur}.  
The latter was based on a specific mapping of the lattice model
to a Coulomb gas.  This is not necessarily inconsistent with
our results,  since the above spectrum is based on the  simplest
closed operator algebra and it contains the main exponents
that are known to be physically meaningful.    It seems
likely that other twisted $\chi$ sectors and/or bosonic $\phi$
sectors could be consistently added to our theory,  however
we have not studied in detail the possibility of obtaining
exactly the partition functions proposed in \cite{ReadSaleur}.

\section{Multi-fractal exponents.}

\subsection{Generalities.}

In order to study multi-fractality in the density of
states,  we add an energy term in the action
corresponding to $H \to H - \CE$:
\beq
\label{m.1}
S_\CE = \int \frac{d^2 x}{2\pi} ~ i \CE 
\( \Psibar_- \Psi_+ + \Psi_- \Psibar_+ 
\) 
\eeq
The density of states operator is then
\beq
\label{m.2}
\rho (x)  =  \Psibar_- \Psi_+ + \Psi_- \Psibar_+ 
\eeq
Multi-fractal properties refer to disorder averages
of q-th moments of $\rho$,  $\bar{\rho^q}$, 
and are simply related to wave-function $\psi$
multi-fractality since $\rho = \psi^\dagger \psi$.    Properly
normalized quantities are
\beq
\label{m.3}
P^{(q)} = 
\frac{ \int   d^2 x  \bar{ \rho (x)^q }}{( \int d^2 x \bar{\rho (x)})^q}
\eeq
where here $\rho$ represents $\langle \rho \rangle$ at fixed disorder. 
At a critical point $P^{(q)}$ scales as
\beq
\label{m.4}
P^{(q)} \sim L^{-\tau_q} 
\eeq
where $L$ is the system size.  The exponents $\tau_q$ are related
to the scaling dimensions of operators as follows:
\beq
\label{m.5}
\tau_q = \Gamma_q - q \Gamma_1 + 2 (q-1)
\eeq
where $\Gamma_q$ is the scaling dimension of $\rho^q$ 
in the effective disorder averaged theory.   

For both the QHT and SQHT,  there is a regime at low $q$ where
$\tau_q$ is quadratic in $q$.  Since $\tau_1 = 0$ and
$\tau_0 = -2$, in this parabolic regime $\tau_q$ is characterized
by a single parameter $\alpha_0$:
\beq
\label{m.7}
\tau_q = (2-\alpha_0 ) q^2 + \alpha_0 q -2 
\eeq
Since the $q \Gamma_1$ term is simply a matter of normalization,
it is meaningful  to define
\beq
\label{m.6} 
\Gammahat_q \equiv \Gamma_q - q \Gamma_1 = (\alpha_0 -2) q (1-q)
\eeq

For the purpose of comparing with numerical simulations, one
can perform the Legendre transformation\cite{Janssen}
\beq
\label{m.8}
f(\alpha) = \alpha q - \tau_q , ~~~~~ \alpha = \frac{d\tau_q}{dq}
\eeq
One finds
$q = (\alpha - \alpha_0)/(2(2-\alpha_0))$,  which leads
to 
\beq
\label{m.9}
f(\alpha) =  - \frac{ (\alpha - \alpha_0)^2}{4 (\alpha_0 -2 )} + 2
\eeq
The parameter $\alpha_0$ determines the typical density
of states $\exp( \bar{\log(\rho)} ) \sim L^{-\alpha_0}$.

A common belief expressed in the literature is that in the
supersymmetric method for disorder averaging,  $q$-th moments
can be calculated in the $N$-copy theory for {\it any} $N$ 
greater than $q$ and one should obtain a result independent of
$N$.   This can be proven by simple manipulations of the path
integral,  and this property is manifested for instance in
the $N$-independence of RG beta functions\cite{networkRG}.
For the QHT (SQHT), it follows from properties of the $gl(N|N)$
($osp(2N|2N)$) BRST symmetries of the theory before and after
disorder averaging.  This ``copy-symmetry'' is analagous to
replica-symmetry.   In the scheme proposed in this paper, 
this copy-symmetry is broken in the first stage of the RG flow
since $gl(N|N)$ is broken to $gl(1|1)$ for the QHT and similarly
for the SQHT, as emphasized in section VII.  Thus,  the $q$-th moments
will depend on the number of copies $N$ one started with.   
Since this is contrary to expectations,  at this stage it must
be viewed as a working hypothesis.   However we can provide
some justification for this copy-symmetry breaking. 
First of all,  the breaking of symmetries in the flow to
a low energy fixed point is a common phenomenon,  and in higher
dimensions is more the rule rather than the exception. 
In the present context one must bear in mind the Mermin-Wagner theorem
which prohibits the spontaneous breaking of symmetries.  
However the breaking of $SU(2) \otimes SU(2)$ to $SU(2)$ based
on spin-charge separation in the $1d$ Hubbard model, which is the
prototype for our RG scheme,  is well-understood and known not
to violate any theorems:  the symmetry is broken in the RG flow
but not spontaneously and there are no Goldstone bosons.   
Secondly, 
in spite of the powerful  map to percolation for the SQHE,
it remains unknown how to obtain the multifractal spectrum
that has been found numerically, i.e. $\Gammahat_q \approx q(1-q)/8$
for continuous $q$ 
from this map.  
(See below.)   Although it may just be a matter of time before
this is eventually understood,  it could instead suggest that some basic
assumptions are incorrect.  
    The fact that we can obtain this result
quite easily in the copy-symmetry broken fixed point 
is a positive indication.    Thirdly,  on the face of it, unbroken copy symmetry
seems incompatible with the phenomenon of multi-fractality termination,
whereas our scheme in fact relies on it to fix $N$,  as we now describe.

For $q$ greater than some critical value $q_c$,  $\tau_q$ is
no longer parabolic.  This phenomenon of 
multi-fractality termination is thought to be  
distinct from the considerations of this paper,
i.e. it is a separate issue unrelated to the RG flow
of the disorder couplings\cite{CCFGM,Caux2,LeD}.   For the QHT,  $2 < q_c < 3$,
whereas for the SQHT, $3 < q_c < 4$.  For the SQHT, 
the map to percolation was used to obtain $\tau_q$ only for the integer values
$q=1,2,3$ in \cite{MirlinSQHT2} and the argument breaks down for $q>3$.  

If one wishes to compute then $\Gammahat_q$ for all of  $q$ continuous and
less than $q_c$  in
a {\it single, fixed theory},  
then this should 
be possible 
in the $N$ copy 
theory with $N$ fixed to be  the largest integer less than $q_c$, 
i.e. $N=2$ for the QHT and $N=3$ for the SQHT.    
We wish to emphasize again that this new approach gives
results that are different from the spectrum of multi-fractal
exponents studied in \cite{Mudry1,Caux2},
and  also differs from the calculation in \cite{networkRG}. 
In the rest of this section we show that these assumptions
appear to give results in very good agreement with numerical work.

\subsection{QHT}

The energy operator corresponds to the $gl(1|1)$
fields
\beq
\label{m.10}
\rho = \bfPhihj{1}{0} + \bfPhihj{0}{1} 
\eeq
We thus first need the $gl(1|1)$ content of 
$\rho^q$ for integer $q$.   Using
\beq
\label{m.11}
\glrep{h}{j} \otimes \( \glrep{1}{0} \oplus \glrep{0}{1} \) 
= \glrep{h+1}{j} \oplus \glrep{h}{j-1} \oplus \glrep{h}{j+1}
\oplus \glrep{h-1}{j} 
\eeq
one sees that the above tensor product involves
$h+j\vert_{\rm new} = h+j \pm 1$ and 
$h-j\vert_{\rm new} = h-j \pm 1$. 
Therefore $\rho^q$ contains the representations $\glrep{h}{j}$
with $-q \leq h+j \leq q$, $-q\leq h-j \leq q$.  
Examining $\Deltahj{h}{j}$ one finds
that the most relevant operator in $\rho^q$ has
$h=0, j=q$.  As explained above, we now set   $N=2=k$, and the field $\bfPhihj{0}{q}$ 
has
dimension 
$\Gamma_q = 2 \Deltahj{0}{q} = q(2-q)/4$, which gives 
\beq
\label{m.12}
\Gammahat_q = \frac{q(1-q)}{4}
\eeq
i.e. $\alpha_0  = 9/4$.
We are implicitely analtyically continuing the $j$ quantum number
to continous $q$.    
This agrees very favorably with the numerical results in
\cite{Klesse}, $\alpha_0 = 2.26\pm .01$,  and
in \cite{MirlinQHT}, $\alpha_0 = 2.260 \pm .003$.  

It needs to be emphasized that the copy-breaking feature of
the RG flow in our scheme implies that $\Gamma_1 = 1/4$ 
in the $N=2$ copy theory is not supposed to equal 
the usual $N=1$ result, i.e. $\Gamma_1 = 0$.

\subsection{SQHT}

\def\osprepNO#1#2{[#1,#2]}

For the SQHT the density operator corresponds to the $osp(2|2)$ field:
\beq
\label{m.13}
\rho = \bfPhi^{osp}_{[0,\inv{2}]}
\eeq
We need the quantum numbers of $\rho^q$ for $q=2,3$.  We have
already considered $q=2$ in eq. (\ref{6.12}).   
Since the  $\eight$ can be viewed as $\osprep{\inv{2}}{\inv{2}} \oplus
\osprep{-\inv{2}}{\inv{2}}$, taking one more tensor product
and using the rules in \cite{Serban} 
one obtains
\beq
\label{m.14}
\osprepNO{0}{\texthalf}\otimes \osprepNO{0}{\texthalf} 
\otimes \osprepNO{0}{\texthalf}
= \osprepNO{0}{{\textstyle \frac{3}{2}}} \oplus 3 \osprepNO{0}{\texthalf}
\oplus 2 \osprepNO{\texthalf}{1}  \oplus 
2 \osprepNO{-\texthalf}{1} \oplus \osprepNO{1}{\texthalf} 
\oplus \osprepNO{-1}{\texthalf}
\eeq
where $\osprepNO{b}{s}$ refers to $\osprep{b}{s}$.  
Examining the conformal dimensions  $\Delta^{osp}_{\osprepNO{b}{s}}$,
one finds that for  $q=1,2,3$ 
 the most relevant operator in $\rho^q$ corresponds
to $\osprep{\frac{q-1}{2}}{\inv{2}}$.   
Fixing  $N=3$, i.e. $k=-6$, as suggested above,   one then has
$\Gamma_q = 2 \Delta^{osp}_{[\frac{q-1}{2},\inv{2}]}  =
q(2-q)/8$
which gives
\beq
\label{m.16}
\Gammahat_q =  \frac{ q(1-q)}{8}
\eeq
In this case  one has $\alpha_0 -2  = 1/8$.  
Again $\Gamma_1 = 1/8$ in the $N=3$ theory is not the same
as the well-known $\Gamma_1 = 1/4$ of the $N=1$ theory
(see section VIIC).

In \cite{MirlinSQHT,MirlinReview} it was found numerically
that $\alpha_0 -2 \approx 1/8$ in the parabolic approximation
(within a few percent).      
However it  was also observed  that in comparison to
the QHT,  for the SQHT there are more marked deviations
from  parabolicity.  If one  takes into account the
non-parabolicity and simply defines $\alpha_0$  
from the  maximum of $f(\alpha)$,  then one obtains
the  result $\alpha_0 -2 = 0.137\pm 0.003$.  
These observed 
deviations from a parabolic regime  could have a number of explanations
in our model.   It could be due to the effects of logarithmic
corrections due to the $\bfPhi_{[8]}$ perturbation.  
It could also be due to the large number of operators in
eq. (\ref{m.14}), where we took only the most relevant.

\section{Localization length exponent.}

In simulations of the network models,  one needs to tune
to a critical point by adjusting a parameter $\lambda$, 
analogous to tuning to the critical probability $p_c = 1/2$ 
in $2D$ classical percolation.   In our description, this
should correspond to a term in the action
\beq
\label{loc.1}
\delta S_\nu = \int \frac{d^2 x}{2\pi}  ~ \lambda \, \CO_\nu (x)
\eeq
for some operator $\CO_\nu$.   If $\CO_\nu$ has scaling
dimension $\Gamma_\nu$, then $\lambda$ has dimension $2-\Gamma_\nu$
and the correlation length diverges as
$\xi_c \sim (\lambda - \lambda_c )^{-\nu} $ with 
$\nu = 1/(2-\Gamma_\nu)$.   

In contrast to the density of states exponents,  we do not have
 arguments based on quantum numbers to identify the field
$\CO_\nu$.    In the SQHT we know that $\Gamma_\nu = 5/4$.
In the $gl(1|1)_2$ embedding in  the 1-copy theory for the SQHT  
the field $\CO_\nu$ thus corresponds to $\bfPhi_{\glrep{2}{1}}$,
which has the explicit form\cite{gl11}:
\beq
\label{loc.2}
\bfPhi_{\glrep{2}{1}} = \mu_{1/2} \bar{\mu}_{1/2}  \, 
\rme^{ i (2\phi^1 - \phi^2 )/\sqrt{2} }
+ \sigma^2_{1/2} \bar{\sigma}_{1/2}^1 \,  \rme^{ i \phi^1 /\sqrt{2} }
\eeq
Above, $\mu_{1/2}$ and $\sigma_{1/2}$ are twist fields
with conformal dimension $-1/8$ and $3/8$ respectively.  

Lacking a first-principles identification of $\CO_\nu$
for the QHT,  we can only give plausible values based on the 
spectrum of dimensions in our model.  
It was understood long ago that one must consider at least $N=2$
copies,  since the exponent describes 
criticality in the conductance.  The latter is
related to a product  of retarded/advanced 2-point
Green functions,  and  one needs separate copies for
retarded verses  advanced. We have already used the  $N=2$
copy exponents to explain the multi-fractality in the density 
of states.   Therefore,  it seems likely that one needs
to consider $N>2$ copies.    
Let us therefore double the number of copies one more time
and consider the $N=4$ theory.  
As explained in  section VII, when  $N>2$ there are
potentially dangerous perturbations corresponding to 
the operators $\bfPhi_{\glrep{0}{2}}$ and $\bfPhi_{\glrep{1}{-1}}$ 
since $\bfPhi_{\hfour{0}}$ is no longer the most relevant operator.
 The possibility of such additional perturbations can  be
 further investigated,  however since this is beyond
 the original scope of this paper,    let us simply assume  this issue is not important,
and consider the theory  $gl(1|1)_4$ plus the logarithmic perturbation
$\bfPhi_{\hfour{0}}$ as in eq. (\ref{glSG}).      
The operator that most closely parallels $\CO_\nu$ for the SQHT 
is  $\bfPhi_{\glrep{N}{N-1}}$  with conformal dimension
\beq
\label{loc.2b}
\Delta_{\glrep{N}{N-1}}  =  \frac{ 2N(N-1) +1 }{2N^2} 
\eeq
in the $N$-copy theory,  and leads to $\nu = N^2/(2N-1)$.     
At $N=4$,  
the operator $\bfPhi_{\glrep{4}{3}}$  has $\Delta = 25/32$ and  has the form 
\beq
\label{loc.3}
\bfPhi_{\glrep{4}{3}} = \mu_{1/4} \bar{\mu}_{3/4} \, 
\rme^{i (4 \phi^1 - 3 \phi^2)/2} 
+ \sigma^2_{1/4} \bar{\sigma}^1_{3/4} \, 
\rme^{i (3 \phi^1 - 2 \phi^2 )/2 } 
\eeq
Here the twist fields $\mu_{1/4}$ and $\sigma_{1/4}^2$ 
have $\Delta = -3/32$ and $5/32$ respectively.  The above fields
are local since $\Deltachi_\lambda = \Deltachi_{1-\lambda}$. 
If one identifies the above operator with $\CO_\nu$, then 
this gives  the exponent $\nu = 16/7 $, which is within $2\%$ 
of the numerical results  $2.35\pm .03$\cite{Huck} and
$2.33 \pm .03$\cite{DHLee}.   (For a survery of the various
methods see \cite{KramerReview}.)  
The value $16/7 \approx 2.29$ is also consistent with 
 the experimental measurement $2.3\pm 0.1$\cite{Koch}.

\section{Conclusions}

In summary,  by carrying out the RG flow in two stages
and using a new form of super spin-charge separation,  we argued that
the disordered Dirac fermion theories for the QHT and SQHT
are described by logarithmic perturbations of the
current algebras $gl(1|1)_N$ and $osp(2|2)_{-2N}$ where
$N$ is the number of copies.  The explicit forms of the
resulting actions were constructed using the recent
results in \cite{gl11}.   We also argued that the
logarithmic perturbations do not modify the exponents of
the current algebra theories,  however they generally lead
to logarithmic corrections to correlation functions. 

The unconventional outcome of our RG scheme is that 
the $N$-copy symmetries implicit in the $gl(N|N)$ symmetry
of the theory before disorder averaging are broken in
the flow to the low energy fixed point in the first stage 
since the $SU(N)_0$ is gapped out in the flow.    This leads
to $N$-dependence of the $q$-th moments,  contrary to 
common expectations.    Although this may seem problematic,
we gave several arguments in favor of it in section VIII,
and showed that it leads to a computation of the   
multi-fractal exponents in the parabolic regime,  which agree
very favorably with known numerical results.     
We also speculated on the localization length exponent for
the QHT, and suggested one needs to consider $N=4$ copies.
After making some plausible  assumptions, we were led to suggest the
value $\nu = 16/7$,  however more investigations of this proposal
are clearly necessary.

The QHT and SQHT are the smallest members of 
two of the main classes of disordered Dirac fermions\cite{BLclass},
and the chiral GUE class was already solved in \cite{Guruswamy}.   
It would be interesting to investigate if the methods
in this paper could be extended to the other classes as well.

\section{Acknowledgments}

I would like to thank the organizers of the 
program {\it Strong fields, Integrability, and Strings}
at the Isaac  Newton Institute for Mathematical Sciences  
during which this work was
begun in July 2007.   
I would also like to thank A. Mirlin for correspondence.

\def\Sh{\Shat}

\section{Appendix A:  Super-current algebras.}

Let $\lie$ denote a finite dimensional super Lie algebra
and $\{ J^a \}$ it generators.   Each generator $J^a$ can 
be assigned a grade $[a]=0$ for bosonic generators and $[a]=1$
for fermionic ones.   The super Lie algebra can be presented as
\beq
\label{A.1}
J^a J^b -  (-)^{[a][b]} J^b J^a = f^{ab}_c J^c 
\eeq
In the super current algebra at level $k$, denoted $\lie_k$, 
the above generators are promoted to fields satisfying the OPE
\beq
\label{A.2}
J^a (z) J^b (0) \sim \frac{k}{z^2} \eta^{ab} + \inv{z} f^{ab}_c J^c (0) 
\eeq
The currents have the mode expansion $J^a (z) = \sum_n J^a_n z^{-n-1}$
and the zero modes $J^a_0$ satisfy eq. (\ref{A.1}).   

An important construction is the Sugawara stress tensor built
on the casimir:
\beq
\label{A.3}
T  =  \kappa   \sum_a J^a J^a 
\eeq
where the coefficient $\kappa$ is fixed by 
the requirement $T(z) J^a (0) \sim  J^a (0)/ z^2  $.  
The conformal scaling dimension of a primary field in
the representation $r$ of $\lie$,   is  given by 
$\kappa C_2$ where $C_2$ is the quadratic casimir 
for $r$.   

Our conventions  for the level $k$ are based on the free field
representations of these algebras in terms of the fields
$\psi_\pm , \beta_\pm$.    Let $\Psi_\pm = (\psi_\pm , \beta_\pm)$ 
denote 2 component fields and $\Psi_\pm^\alpha$, $\alpha = 1,2,..,N$
the $N$-copy version.   Let us arrange all these fields into
$2N$ component fields $\Psi_\pm^a$,  $a=1,..,2N$
and let $[a]$ be the grade.   A complete basis of currents
is defined  in eq. (\ref{4.3}).   OPE's can be readily computed
from the OPE's of the $\Psi_\pm^a$ in eq. (\ref{4.2})
and will serve as our defining relations for the level $k=1$ 
super current algebra.   For instance, 
\beq
\label{A.4}
H^{ab} (z) H^{cd} (0) \sim 
\sim \frac{k}{z^2} (-)^{[b]+1} \delta^{bc} \delta^{ad} 
+ \inv{z} \(  (-)^{[b]+1} \delta^{bc} \, H^{ad}  
+ (-)^{[a]([b]+[c]) + [b][c] } \delta^{ad} \, H^{cb}   \)
\eeq
where $k=1$.   There are similar  relations of the form
$H(z)  J_\pm (0)  \sim   J_\pm /z  $ and
$J_+ (z) J_- (0) \sim k/z^2  + H / z  $.  
We take the above OPE's as the defining relations of
$osp(2N|2N)_k$ for general $k$.   The $H^{ab}$ form a closed
subalgebra which defines $gl(N|N)_k$.  

Generally the Lie super-algebra $\lie$ can be decomposed into 
its bosonic generators $\lie^{(0)}$ and fermionic generators $\lie^{(1)}$,
$\lie = \lie^{(0)}  \oplus \lie^{(1)}$.  The bosonic generators form a closed
subalgebra,  and the fermionic generators fall into representations
of $\lie^{(0)}$.   For $osp(2N|2N)$, $\lie^{(0)}= so(2N) \oplus sp(2N)$
and $\lie^{(1)}$ corresponds to the representation $(2N,2N)$.   
The dimension of $osp(2N|2N)$ is thus $8N^2$ and its rank is $2N$.  

The algebra $gl(N|N)$ is $psl(N|N) \oplus u(1) \oplus  u(1)$  where
$psl(N|N)$ is denoted $A(N-1,N-1)/Z$ in \cite{Dictionary}.  
The bosonic subalgebra is $\lie^{(0)} = sl(N) \oplus sl(N) \oplus u(1)
\oplus u(1)$.  It has dimension $4N^2$ and it's rank is $2N$.   

For the  current algebras $gl(1|1)_k$ and $osp(2|2)_k$ we adopt
a more specific notation.   Again our conventions for the levels
are based on the $N=1$ copy of the fields $\psi_\pm , \beta_\pm$.
Define:
\barray
\nonumber
H &=& \psi_+ \psi_- , ~~~~~J= \beta_+ \beta_- , ~~~~~J_\pm = \beta_\mp^2 
\\ \label{A.5}
S_\pm &=& \pm \psi_\pm \beta_\mp , ~~~~~ \Shat_\pm = \psi_\mp \beta_\mp
\earray
These currents satisfy the OPE's
\begin{eqnarray}
\nonumber
J(z) J(0) &\sim& - \frac{k}{z^2} ,  
~~~~~~~~~~~~~~
~H(z) H(0) \sim \frac{k}{z^2}
\\ \nonumber
J(z) J_\pm (0) &\sim&  \pm \frac{2}{z} ~ J_\pm
, ~~~~~~~~~~
J_+(z) J_- (0) \sim \frac{2k}{z^2} - \frac{4}{z} J
\\ \nonumber
J(z) S_\pm (0) &\sim&  \pm \inv{z} S_\pm   ,  
~~~~~~~~~~ J(z) \Sh_\pm (0) \sim
\pm \inv{z} \Sh_\pm
\\ \nonumber
H(z) S_\pm (0) &\sim&  \pm \inv{z} S_\pm  , ~~~~~~~~~~
H(z) \Sh_\pm (0) \sim \mp \inv{z} \Sh_\pm
\\ \label{3.8}
J_\pm (z) S_\mp (0) &\sim& \frac{2}{z} \Sh_\pm , 
~~~~~~~~~~~~~~~
J_\pm (z) \Sh_\mp (0) \sim - \frac{2}{z} S_\pm
\\
\nonumber
&&
S_\pm (z) \Sh_\pm (0) \sim \pm \inv{z} J_\pm
\\ \nonumber
&&
S_+ (z) S_- (0) \sim  \frac{k}{z^2} + \inv{z} (H-J)
\\ \nonumber
&&
\Sh_+ (z) \Sh_- (0) \sim - \frac{k}{z^2} + \inv{z} (H+ J)
\end{eqnarray}
Since $sp(2) = su(2)$,  the bosonic subalgebra generated by
$H, J , J_\pm$ corresponds to $su(2) \oplus u(1)$ where
the $J, J_\pm$ are $su(2)$ generators.   Rescaling
$J \to 2J $,  $J_\pm \to \pm 2 \sqrt{2} J_\pm$ one finds
they satisfy the $su(2)$ current algebra at level $-k/2$.  
Thus, $osp(2|2)_k$ has an $su(2)_{-k/2}$ subalgebra.  

Our conventions for $gl(1|1)$ are taken from the subalgebra
of $osp(2|2)_k$ generated by $S_\pm , H, J$ and are presented in
eq. (\ref{5.2}).   The stress tensors and scaling
dimensions of primary fields are given in the main body of
the paper.  

Note that letting $H\to J$, $J\to H$,  $S_\pm \to \pm S_\pm$, the
new currents satisfy $gl(1|1)_{-k}$.  Thus there exists 
an automorphism of $gl(1|1)_k$ which flips the sign of $k$,
and  was used in\cite{gl11}.

\vspace{.4cm}
\end{document}